# CLASSIFYING $K$-GEODETIC GRAPHS VIA COMBINATORIAL DESIGNS AND DISTANCE-REGULARITY

**Frasser C.E.**
*Ph.D. in Engineering Sciences, Odessa National Polytechnic University, Ukraine*

**Abstract**

The classification of $K$-geodetic graphs is a fundamental problem in graph theory concerning the number of shortest paths connecting pairs of vertices. A graph is called $K$-geodetic if every pair of distinct vertices is joined by at most $K$ geodesics, where a geodesic is a shortest path between two vertices. The cases $K = 1, 2, 3$ correspond, respectively, to geodetic, bigeodetic, and trigeodetic graphs. Despite substantial progress in the study of their structural properties, a comprehensive classification of these graph families remains elusive. This paper develops a research framework for investigating the classification problems by combining metric graph theory, distance-regular graphs, algebraic graph theory, and combinatorial design theory. Particular attention is given to distance-regular graphs, for which the number of geodesics between vertices at a fixed distance is determined by the intersection numbers of the graph. This property converts the $K$-geodetic condition into a finite collection of arithmetic inequalities. We establish the resulting necessary conditions for $K = 1, 2, 3$, discuss representative examples, and distinguish these conditions from the stronger requirements needed to prove the existence and classification of graphs with a given intersection array. The paper also investigates the potential role of balanced incomplete block designs (BIBDs) in constructing and organizing families of $K$-geodetic graphs. Finally, it proposes a computational research methodology based on constraint satisfaction, graph-isomorphism testing, exhaustive enumeration, and machine-assisted conjecture generation. The central thesis is that a general classification is more likely to emerge from a hierarchy of structural and algebraic classifications than from a single universal characterization. Distance-regular graphs and BIBDs provide a particularly promising setting in which to develop and test this approach.

## 1. Introduction

### 1.1. Background and Motivation

Graph theory provides a mathematical language for studying networks whose vertices represent objects and whose edges represent relationships between them. Among the many properties used to distinguish graph families, distance is particularly important because it measures how efficiently two vertices can communicate through the network.

Let $G = (V, E)$ be a finite, simple, connected graph. For vertices $u, v \in V$, the distance $d(u, v)$ is the minimum number of edges in a path connecting $u$ and $v$. A path whose length equals $d(u, v)$ is called a *geodesic*.

In general, two vertices may be connected by several geodesics of the same minimum length. The number of such paths reflects the degree of ambiguity in shortest-route communication. This observation motivates the definition of $K$-geodetic graphs.

**Definition 1.1.** Let $K$ be a positive integer. A connected graph $G$ is called $K$-geodetic if every pair of distinct vertices is connected by at most $K$ distinct geodesics.

The terminology for the first three cases is:

- $K = 1$: geodetic graphs, in which every pair of vertices has a unique shortest path.
- $K = 2$: bigeodetic graphs, in which every pair has at most two shortest paths.
- $K = 3$: trigeodetic graphs, in which every pair has at most three shortest paths.

The definition concerns shortest paths only. A graph may contain many paths between two vertices without violating the $K$-geodetic condition, provided that no more than $K$ of those paths are shortest.

*The classification problem asks which graphs satisfy this restriction, how they can be constructed, and whether their structures can be described by a finite collection of theorems or parametrized families.*

The historical motivation goes back to Øystein Ore's 1962 treatment of geodetic graphs. Subsequent research has revealed connections with strongly regular graphs, finite geometry, combinatorial designs, and distance-regular graphs. These connections suggest that the classification problem should be approached through several complementary mathematical frameworks.

### 1.2. The central Difficulties

**1. Global vs. Local Constraints.** This difficulty arises from the interaction between a local graph structure and a global metric condition. Most easily classifiable graph families are defined by local properties (e.g., forbidden subgraphs or minors). *Geodeticity* is a global metric property dependent on the entire graph's shortest-path structure. A single edge change can alter geodesics across the whole graph. Consequently, a condition involving the number of shortest paths cannot generally be verified by inspecting only small neighborhoods.

**2. Links to Deep Open Problems.** A second obstacle is the existence problem. Some geodetic graphs of diameter two coincide with strongly regular graphs with parameter $\mu = 1$. This family includes the Petersen graph and the Hoffman-Singleton graph. Whether infinitely many such graphs exist is a legendary open problem. Even when a proposed set of numerical parameters satisfies all elementary counting constraints, there may be no constructible graph satisfying those parameters. A complete classification must therefore answer two distinct questions:

a. **Existence:** Does a graph with the proposed structure exist?

b. **Classification:** If it exists, what are all graphs with that structure, up to graph isomorphism?

The distinction is essential when algebraic parameters are used to organize possible graph families.

**3. Combinatorial Design Theory**. Modern research shows that $K$-geodetic graphs can be constructed using BIBDs [1]. Classifying these graphs requires a full classification of BIBDs, which is itself intractable due to immense structural variation as the number of elements grows.

**4. Structural Explosion for $K = 2$ and $K = 3$.** While geodetic graphs cannot contain even cycles with matching opposite distances, bigeodetic and trigeodetic graphs allow ladder-like configurations and even cycles, defying a unified framework.

### 1.3. Research Objectives

The objective of this paper is to develop a mathematically controlled framework for the general classification of $K$-geodetic graphs, emphasizing the following goals.

First, we formulate the $K$-geodetic condition in metric and combinatorial terms. Second, we investigate distance-regular graphs, where intersection numbers yield an explicit formula for the number of geodesics between vertices at a given distance. Third, we examine the role that combinatorial designs may play in constructing and classifying suitable graph families. Fourth, we outline how computational methods and artificial intelligence can assist in discovering examples and testing conjectures without replacing mathematical proof.

The paper distinguishes established mathematical facts, necessary conditions, illustrative examples, and proposed research directions. This distinction is important because a useful classification framework should not be confused with a completed classification theorem.

### 1.4. Organization

Section 2 introduces the basic definitions and counting principles. Section 3 develops the theory of distance-regular graphs and derives the central geodesic-counting formula. Section 4 analyzes the cases $K = 1, 2, 3$. Section 5 discusses strongly regular graphs and diameter-two restrictions. Section 6 analyzes in detail the use of BIBDs as algebraic seeds for the general classification problem. Section 7 explains why BIBDs and distance-regular graphs cannot solve the entire problem by themselves, while outlining the key differences and trade-offs. Section 8 presents a computational classification strategy, including the potential role of artificial intelligence. Section 9 formulates research questions and conjectural directions. Section 10 discusses limitations and implications, followed by the conclusions and references.

## 2. Fundamental Definitions and Combinatorial Principles

### 2.1. Geodesics and Geodesic multiplicity

Let $G = (V, E)$ be a finite, simple, connected graph. For $u, v \in V$, define

$$d(u, v) = \min\{|P|: P \text{ is a path from } u \text{ to } v\},$$

where $|P|$ denotes the number of edges in the path $P$.

Let $\sigma(u, v)$ denote the number of distinct geodesics connecting $u$ and $v$. Then

$$G \text{ is } K - \text{Geodetic} \Leftrightarrow \sigma(u, v) \leq K \text{ for all } u \neq v.$$

For adjacent vertices, the edge joining them is their unique shortest path in a simple graph. Thus, the condition is equivalent to imposing the bound only on nonadjacent pairs, as is customary in the literature.

A useful way to express the condition is through the *maximum geodesic multiplicity*:

$$\gamma(G) = \max_{u \neq v} \sigma(u, v),$$

Consequently,

$$G \text{ is } K - \text{Geodetic} \Leftrightarrow \gamma(G) \leq K.$$

This formulation gives a natural hierarchy:

$$\mathcal{G}_1 \subseteq \mathcal{G}_2 \subseteq \mathcal{G}_3 \subseteq \cdots$$

where $\mathcal{G}_K$ denotes the class of connected $K$-geodetic graphs.

The inclusions are immediate from the definition. However, they do not imply that the difference between successive classes is structurally simple.

### 2.2. Why Shortest-path Multiplicity is a Global Invariant

Consider a pair of vertices $u, v$ at distance $d$. Their shortest paths are precisely those paths of length $d$. A path of length $d + 1$ or greater does not contribute to $\sigma(u, v)$.

This distinction matters because the number of paths in a graph is not the same as its geodesic multiplicity. For example, a graph can have many distinct paths between two vertices but remain geodetic if exactly one of those paths is shortest. Conversely, the presence of only a few paths does not guarantee that the graph is geodetic: two paths of equal minimum length are enough to violate the $K = 1$ condition.

The relevant property is therefore not connectivity alone, but the interaction between connectivity and distance.

### 2.3. A Basic Example: Cycles

Let $C_n$ denote the cycle on $n$ vertices.

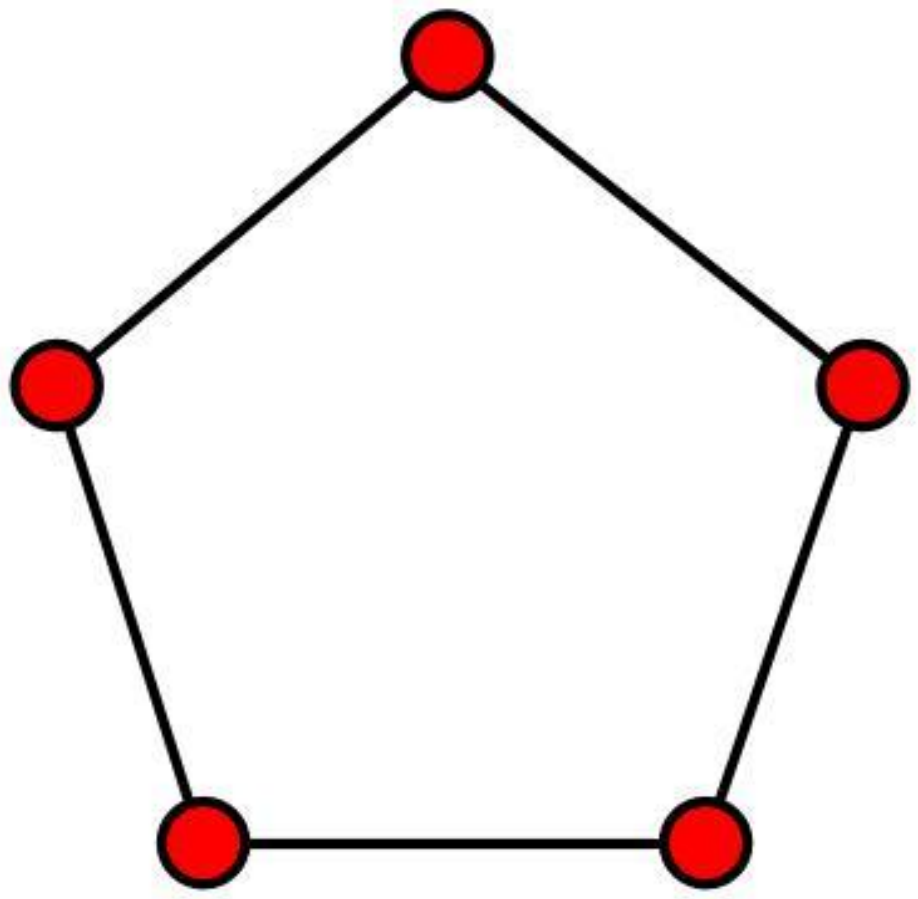

Fig. 1. Odd Cycles

For an odd cycle, the two arcs connecting any pair of vertices have different lengths. The shorter arc is therefore unique. In $C_{2m+1}$, every pair of vertices has a unique shortest path. Hence the cycle is geodetic.

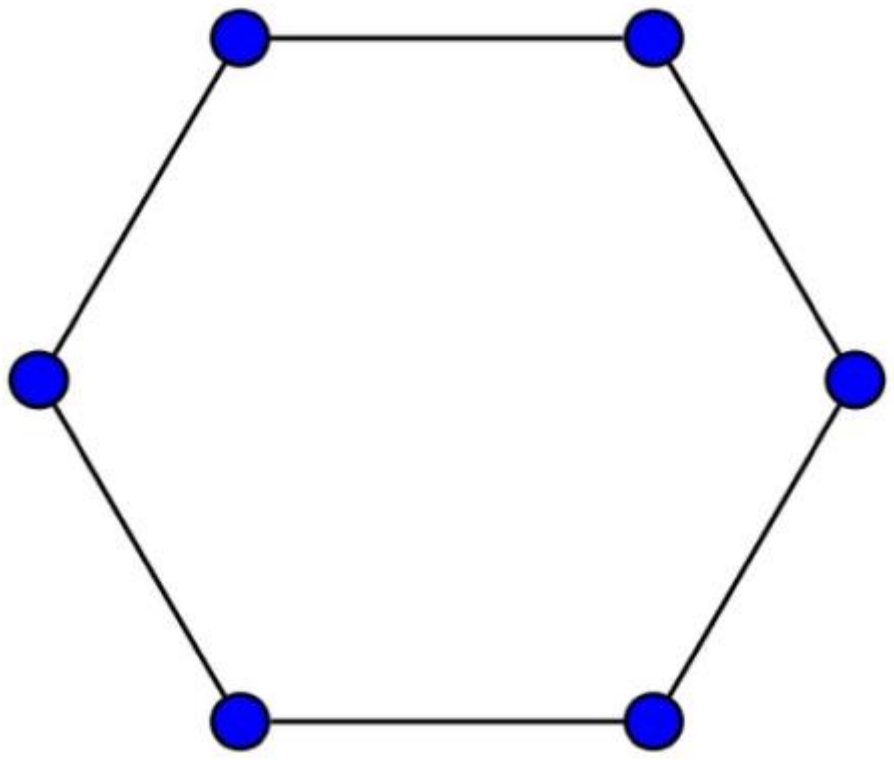

Fig. 2. Even Cycles

For an even cycle, every pair of opposite vertices is connected by two arcs of equal length. These are both shortest paths, so the cycle is not geodetic. Therefore, in $C_{2m}$, opposite vertices have two shortest paths. Hence the cycle is bigeodetic.

Thus, for cycles

$$\gamma(G) = \begin{cases} 1, n \text{ odd}, \\ 2, n \text{ even}. \end{cases}$$

This example illustrates an important principle: *a small change in the global geometry of a graph can change its geodesic multiplicity even when its local structure remains nearly identical*.

## 3. Distance-Regular Graphs as an Algebraic Framework

### 3.1. Definition of Distance-regularity

The general classification problem involves graphs with widely varying local structures. A natural strategy is to restrict attention to a family in which distances obey strong uniformity conditions.

A connected graph $G$ of diameter $d$ is *distance-regular* if, for every pair of vertices $x, y$ at distance $i$, the numbers of neighbors of $y$ at distances $i-1$, $i$, and $i+1$ from $x$ depend only on $i$, not on the particular vertices chosen. These numbers are denoted by $c_i,\ a_i,\ b_i$, respectively.

For a fixed vertex $x$, let

$$\Gamma_i(x) = \{y \in V : d(x, y) = i\}$$

be the set of vertices at distance $i$ from $x$.

If $y \in \Gamma_i(x)$, then:

- $c_i$ counts the neighbors of $y$ in $\Gamma_{i-1}(x)$;
- $a_i$ counts the neighbors of $y$ in $\Gamma_i(x)$;
- $b_i$ counts the neighbors of $y$ in $\Gamma_{i+1}(x)$.

The intersection numbers satisfy

$$a_i + \ b_i + c_i = k,$$

where $k$ is the common vertex degree. The boundary conditions include

$$c_1 = 1,\ a_0 = 0,\ b_d = 0.$$

The sequence of intersection numbers can be recorded in an intersection array:

$$\iota(G) = \{\ b_0,\ b_1, \dots,\ b_{d-1}; c_1, c_2, \dots, c_d\}.$$

This array captures substantial information about the graph's distance structure. In particular, it determines the number of vertices at each distance from any fixed vertex. It does not, by itself, guarantee that every graph satisfying the numerical constraints exists or that only one constructible graph satisfies them.

### 3.2. Counting Shortest Paths using Intersection Numbers

The central observation for the present investigation is that distance-regularity determines geodesic multiplicity.

**Theorem 3.1.** Let $G$ be a distance-regular graph. If $x, y \in V(G)$ satisfy $d(x,y) = i$, then the number of geodesics connecting $x$ and $y$ is

$$\sigma(x,y) = \prod_{j=1}^{i} c_j.$$

*Proof.* We proceed by induction on $i = d(x,y)$.

If $i = 1$, the vertices are adjacent. Their shortest path is the edge $xy$, so

$$\sigma(x,y) = 1 = c_1.$$

Now suppose the formula holds for pairs at distance $i - 1$, and consider vertices $x, y$ with $d(x,y) = i$.

Every geodesic from $x$ to $y$ must reach $y$ through a neighbor $z$ of $y$ satisfying

$$d(x,z) = i - 1.$$

By distance-regularity, there are exactly $c_i$ such neighbors $z$. For each such neighbor, the induction hypothesis gives exactly

$$\prod_{j=1}^{i-1} c_j$$

geodesics from $x$ to $z$. Appending the edge $zy$ produces a geodesic from $x$ to $y$. Different choices of $z$, or different geodesics to the same $z$, produce distinct geodesics.

It follows that

$$\sigma(x,y) = c_i \prod_{j=1}^{i-1} c_j = \prod_{j=1}^{i} c_j$$

The result follows by induction. □

### 3.3. The Resulting Classification Criterion

Because the product in Theorem 3.1 depends only on the distance $i$, every pair of vertices at the same distance has the same number of geodesics.

Consequently, a distance-regular graph of diameter $d$ is $K$-geodetic if and only if

$$\prod_{j=1}^{i} c_j \leq K \text{ for every } 1 \leq i \leq d$$

This is the principal algebraic reduction in the paper.

The general problem asks us to inspect every pair of vertices and count their shortest paths. Within the distance-regular class, that potentially enormous collection of checks is replaced by $d$ inequalities involving positive integers.

For example, suppose a distance-regular graph has intersection parameters

$$c_1 = 1,\ c_2 = 1,\ c_3 = 2.$$

The numbers of geodesics between pairs at distances one, two, and three are, respectively,

$$1,\quad 1,\quad 2.$$

If the graph exists with these parameters, it satisfies the bigeodetic condition. Its intersection array must, of course, satisfy all other requirements for a distance-regular graph to exist.

The important distinction is that the product criterion tests the $K$-geodetic property of a distance-regular graph; it does not independently prove the existence of a graph with an arbitrarily proposed intersection array.

### 3.4. Monotonicity of the Intersection Numbers

For distance-regular graphs, the sequence

$$c_1, c_2, \dots, c_d$$

is nondecreasing. Since $c_1 = 1$, all subsequent terms are positive integers satisfying

$$1 = c_1 \leq c_2 \leq \cdots \leq c_d$$

This elementary-looking restriction has significant consequences. Once some $c_i$ exceeds one, all later $c_j$ must be at least as large.

Therefore, the product

$$c_1 c_2 \dots c_i$$

cannot decrease as $i$ increases. A small upper bound on the product severely restricts the allowable intersection numbers.

This observation provides the starting point for analyzing the three principal cases.

## 4. The Cases $K = 1$, $K = 2$, and $K = 3$

### 4.1. Geodetic Distance-regular Graphs: $K = 1$

For a distance-regular graph to be geodetic, Theorem 3.1 requires

$$\prod_{j=1}^{i} c_j \leq 1 \quad (1 \leq i \leq d).$$

Every $c_j$ is a positive integer. Hence every factor must equal one:

$$c_1 = c_2 = \cdots = c_d = 1.$$

Conversely, this condition implies that every pair of vertices has exactly one geodesic. Thus, among distance-regular graphs, the geodetic property is equivalent to the condition that all the $c_i$ equal one.

This gives a necessary and sufficient criterion within the specified class, but it is not yet a complete structural classification. The next task is to determine which intersection arrays with these values are realizable.

#### 4.1.1. Examples

Odd cycles provide an infinite family of geodetic distance-regular graphs. In an odd cycle $C_{2m+1}$, the intersection numbers $c_i$ equal one at all distances up to the diameter $m$.

Another important family comes from complete graphs, which have diameter one and a unique geodesic between every pair of vertices.

Graphs of diameter two provide a different source of examples. A strongly regular graph with parameters $(v, k, \lambda, \mu)$ has the property that every pair of nonadjacent vertices has exactly $\mu$ common neighbors. If $\mu = 1$, every nonadjacent pair has a unique path of length two. Since every adjacent pair has a unique geodesic of length one, the graph is geodetic.

Such graphs include the Petersen graph and the Hoffman-Singleton graph. These examples are particularly significant because they also belong to the family of Moore graphs, although the condition $\mu = 1$ alone does not imply that a strongly regular graph is a Moore graph.

#### 4.1.2. The Classification Question

The distance-regular problem for $K = 1$ can therefore be formulated as follows:

*Determine all realizable intersection arrays satisfying $c_i = 1$ for every $i$, and classify all distance-regular graphs realizing those arrays.*

This is a narrower problem than classifying every geodetic graph. It is also a mathematically precise first step toward a broader classification.

### 4.2. Bigeodetic Distance-regular Graphs: $K = 2$

The bigeodetic condition is

$$\prod_{j=1}^{i} c_j \leq 2 \quad (1 \leq i \leq d).$$

Since $c_1 = 1$ and the $c_i$ are nondecreasing positive integers, the possibilities are sharply restricted.

If every $c_i = 1$, the graph is geodetic and hence bigeodetic.

Otherwise, the first value exceeding one must equal two. If $c_i = 2$ for some $i < d$, then $c_{i+1} \geq 2$, giving

$$\prod_{j=1}^{i+1} c_j \geq 4,$$

which violates the bigeodetic condition.

Therefore, the only non-geodetic possibility is

$$c_1 = c_2 = \cdots = c_{d-1} = 1, \; c_d = 2.$$

Together with the all-ones case, this is a complete description of the *allowable* $c_i$-sequences for bigeodetic distance-regular graphs.

#### 4.2.1. Even Cycles

The even cycles $C_{2m}$ illustrate the second possibility. Their diameter is $m$, and

$$c_1 = c_2 = \cdots = c_{m-1} = 1, \; c_m = 2.$$

Every pair at distance less than $m$ has a unique geodesic. Opposite vertices, which are at distance $m$, have two shortest paths.

The even cycles are therefore bigeodetic but not geodetic.

#### 4.2.2. What Remains to Be Proved

It would be incorrect to conclude from the sequence restriction that even cycles are the only non-geodetic bigeodetic distance-regular graphs.

The restriction describes the possible values of the $c_i$, not all the other intersection numbers. Further conditions involving $a_i$, $b_i$, the spectrum, and the existence of a graph with the proposed parameters must also be considered.

The appropriate classification problem is therefore to determine all realizable intersection arrays of the form

$$\{ b_0, b_1, \dots, b_{d-1}; 1, 1, \dots, 1, 2\},$$

along with the arrays having all $c_i = 1$.

### 4.3. Trigeodetic Distance-regular Graphs: $K = 3$

For $K = 3$, we require

$$\prod_{j=1}^{i} c_j \le 3 \ (1 \le i \le d).$$

Again, monotonicity and integrality restrict the possible sequences.

If all $c_i = 1$, the graph is geodetic.

If the first value greater than one is two, it must occur at the final distance; otherwise, the product becomes at least four at the next distance.

Alternatively, the first value greater than one can be three, but it too must occur only at the final distance. We obtain the following possibilities:

$$c_1 = \cdots = c_d = 1,$$

$$c_1 = \cdots = c_{d-1} = 1, \ c_d = 2,$$

$$c_1 = \cdots = c_{d-1} = 1, \ c_d = 3.$$

These conditions are necessary and sufficient for the product inequalities within the distance-regular framework.

#### 4.3.1. Interpretation of the Final Case

Suppose that

$$c_1 = \cdots = c_{d-1} = 1, \ c_d = 3.$$

The geodesic multiplicity between vertices at distances less than $d$ is one. At distance $d$, the multiplicity becomes three.

In other words, the distance structure permits unique shortest paths up to the penultimate distance, while the final layer introduces exactly three geodesics.

This is a precise algebraic description of the permitted path multiplicity. It does not, by itself, guarantee that a graph with such parameters exists, nor does it establish that the graph must arise from a particular covering construction.

Any proposed example must be checked against the full set of distance-regularity conditions and the remaining intersection parameters.

### 4.4. Comparison of the Three Cases

| **Property** | $\boldsymbol{K = 1}$ | $\boldsymbol{K = 2}$ | $\boldsymbol{K = 3}$ |
|---|---|---|---|
| Geodesics permitted | At most one | At most two | At most three |
| Possible $c_i$-sequence | All ones | All ones, or final value two | All ones, or final value two or three |
| Example | Odd cycles | Even cycles | Includes all $K = 1$ and $K = 2$ examples |
| Main algebraic restriction | $c_i = 1$ | Product at most two | Product at most three |
| Remaining task | Classify realizable arrays | Classify realizable arrays | Classify realizable arrays |

The central conclusion is that the distance-regular case admits a clean reduction for all three values of $K$. The challenge is then shifted from counting geodesics to classifying realizable intersection arrays and the graphs that realize them.

## 5. Strongly Regular Graphs and Diameter-Two Classification

### 5.1. Strongly Regular Graphs

A graph is strongly regular with parameters $(v, k, \lambda, \mu)$ if:

1. It has $v$ vertices.
2. Every vertex has degree $k$.
3. Every pair of adjacent vertices has exactly $\lambda$ common neighbors.
4. Every pair of nonadjacent vertices has exactly $\mu$ common neighbors.

A connected, noncomplete strongly regular graph is distance-regular with diameter two. Its intersection array is

$$\{k, k - \lambda - 1;\ 1, \mu\}.$$

To see why, consider adjacent vertices $x, y$. The neighbors of $y$ consist of $\lambda$ vertices adjacent to $x$, one vertex $x$, and the remaining $k - \lambda - 1$ vertices at distance two from $x$. This gives

$$a_1 = \lambda,\ b_1 = k - \lambda - 1.$$

For nonadjacent vertices $x, y$, their common neighbors are exactly the neighbors of $y$ at distance one from $x$, so $c_2 = \mu$. The number of geodesics between nonadjacent vertices is therefore precisely $\mu$. It follows immediately that:

**Theorem 5.1.** A connected, noncomplete strongly regular graph is $K$-geodetic if and only if

$$\mu \leq K.$$

*Proof.* Adjacent pairs have a unique geodesic of length one. Every nonadjacent pair has distance two and has exactly $\mu$ common neighbors. Each common neighbor yields a distinct geodesic of length two, and every such geodesic arises in this way.

Hence the maximum geodesic multiplicity is $\mu$, proving the result. □

This theorem illustrates the value of algebraic structure. A condition involving all shortest paths becomes a single inequality on a strongly regular graph parameter.

### 5.2. The Special Case $\mu = 1$

When $\mu = 1$, every nonadjacent pair has exactly one common neighbor. Therefore, the graph is geodetic.

This family includes several famous graphs:

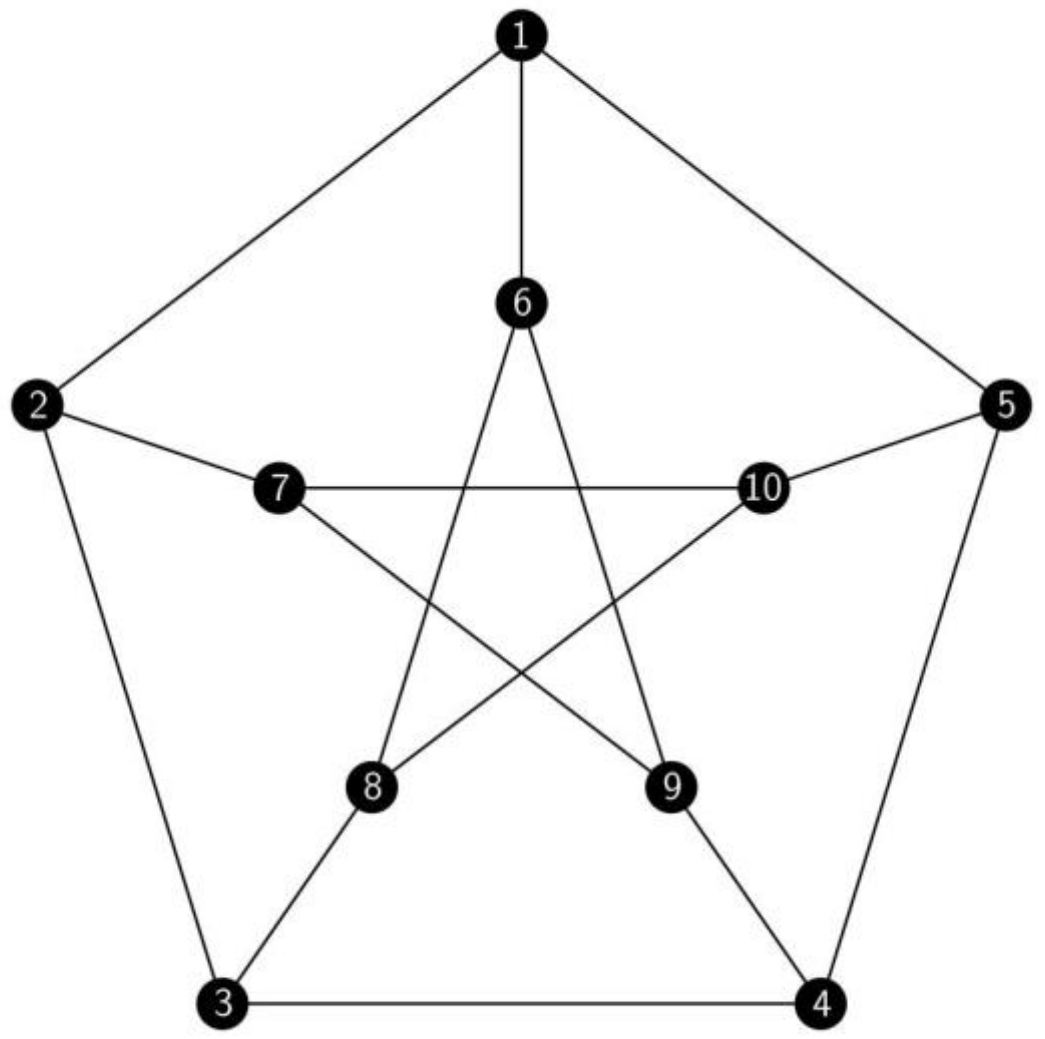

Fig. 3. The Petersen Graph

Petersen graph: Parameters $(10, 3, 0, 1)$. It is strongly regular, distance-regular, and geodetic.

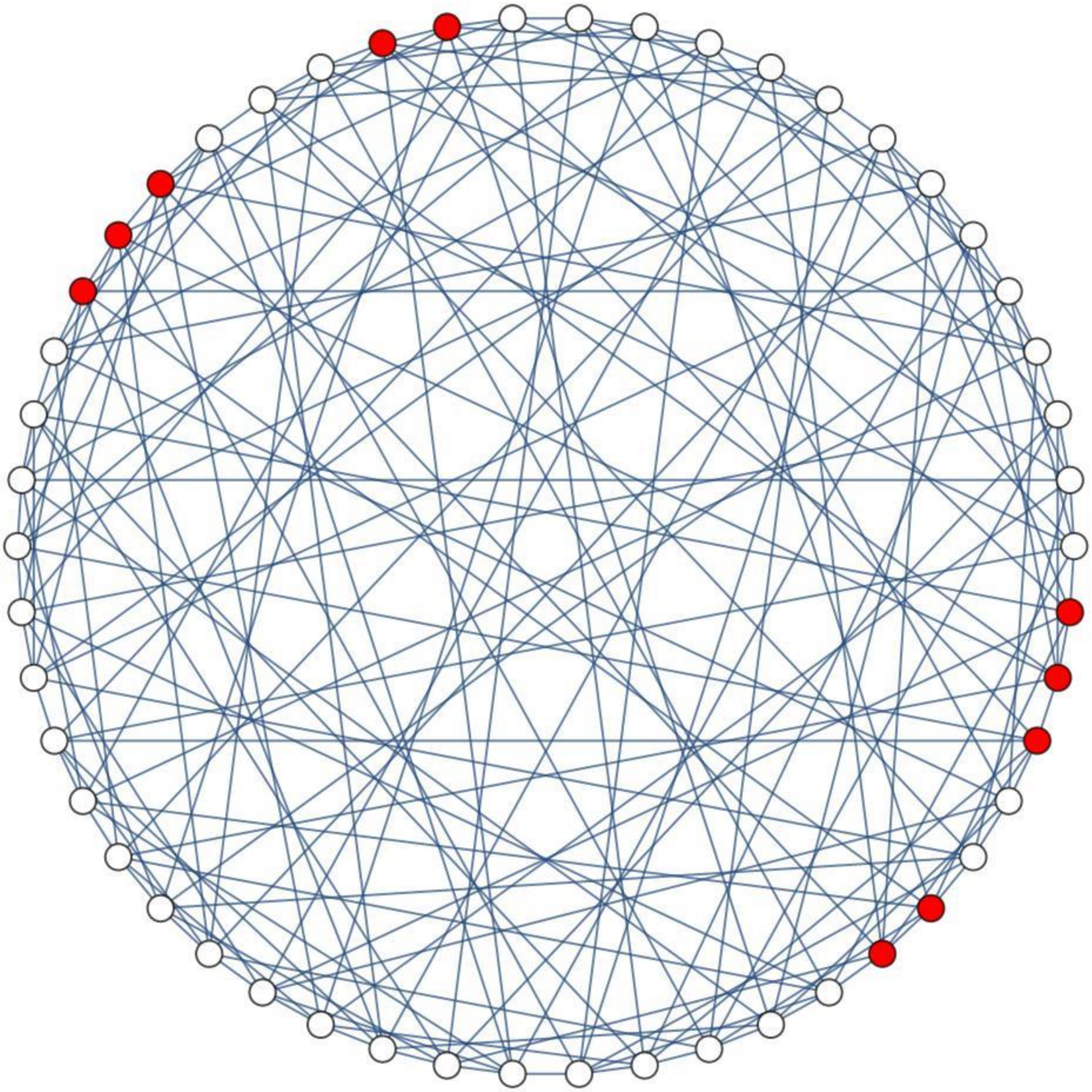

Fig. 4. The Hoffman-Singleton Graph

Hoffman–Singleton graph: Parameters $(50, 7, 0, 1)$. It is another celebrated example of a geodetic strongly regular graph. In Fig. 4, the 10 red vertices highlight a specific subset of nodes that form a *graphical design*. In graph theory, a graphical design is a concept analogous to a *spherical design* in geometry. It represents a subset of vertices that behave like a highly balanced, evenly distributed sampling of the entire graph. Specifically, this particular set of 10 red vertices satisfy the following properties:

**Mathematical Integration:** They perfectly integrate (or average out) the first 49 eigenvectors of the graph's Laplacian matrix in a suitable basis.

**Substructure:** If you look at those 10 red vertices and the connections between them, they form a *3-regular induced graph* (meaning every red vertex connects to exactly three other red vertices).

The color coding is used purely to provide a clear visual representation of how this subset is geometrically embedded within the highly symmetrical 50-vertex structure.

Both the Petersen Graph and the Hoffman-Singleton graph are Moore graphs: they have diameter two and girth five. Their importance to geodetic graph theory comes from the fact that their nonadjacent vertex pairs have unique common neighbors.

However, the equation $\mu = 1$ is more general than the Moore graph condition. A strongly regular graph can have $\mu = 1$ and $\lambda > 0$, in which case it is still geodetic even though it contains triangles and does not have girth five.

This distinction prevents a common overstatement: *geodetic strongly regular graphs are not necessarily restricted to the Moore graph examples*.

### 5.3. The Moore Graph Connection

A Moore graph of diameter two has degree $k$, diameter two, and girth five. Every vertex has $k$ neighbors, and the absence of triangles and four-cycles forces all vertices within distance two of a fixed vertex to be distinct.

Consequently, its number of vertices must be

$$v = 1 + k + k(k - 1) = k^2 + 1.$$

Equivalently, its strongly regular parameters are

$$(v, k, \lambda, \mu) = (k^2 + 1, k, 0, 1).$$

The classical degree restrictions imply that a Moore graph of diameter two can have degree 2, 3, 7, or 57 with the degree-57 case remaining unresolved.

The known examples are the cycle $C_5$, the Petersen graph, and the Hoffman–Singleton graph. The possible degree-57 graph has not been established to exist.

*This is a useful illustration of the difficulty of classification: even strong algebraic restrictions and precise counting arguments may leave an existence problem unresolved*.

Nevertheless, the Moore graph problem is a special problem within algebraic graph theory, not a complete description of all geodetic graphs.

Finally, it is worthy to mention that there exist the entire infinite families of geodetic graphs homeomorphic to a complete graph $\mathcal{K}_n$ (that is, the graphs $\mathcal{K}_n^i$) and also the ones homeomorphic to the Petersen graph and the Hoffman-Singleton graph [2]. Additionally, the generation and enumeration of bigeodetic and trigeodetic graphs homeomorphic to any geodetic graph (including complete graphs, the Petersen graph, and the Hoffman-Singleton graph) remain open problems, pending the verification of their existence.

### 5.4. Implications for the $K$-geodetic Problem

Theorem 5.1 shows that diameter-two strongly regular graphs provide an immediate classification route for each fixed $K$.

For $K = 1$, the relevant parameter condition is $\mu = 1$.

For $K = 2$, the possible strongly regular parameter values satisfy $\mu \in \{1, 2\}$.

For $K = 3$, they satisfy $\mu \in \{1, 2, 3\}$.

These are necessary and sufficient conditions for the $K$-geodetic property within the strongly regular class. They do not classify all strongly regular graphs with those parameters, nor do they cover graphs of larger diameter.

The distinction is important because a classification theorem should identify the precise class under consideration and state which part of the problem has been solved.

## 6. Balanced Incomplete Block Designs as Algebraic Seeds

### 6.1. The Motivation for using Block Designs

The study of $K$-geodetic graphs is not confined to distance-regular graphs. General graphs may lack the uniform intersection numbers required in Section 3, making the algebraic reduction unavailable.

A complementary strategy is to construct graphs from combinatorial designs.

A balanced incomplete block design (BIBD) with parameters

$$(v, b, r, k, \lambda)$$

consists of:

- a set $X$ of $v$ points;
- a collection of $b$ blocks;
- exactly $k$ points in each block;
- exactly $r$ blocks containing each point;
- exactly $\lambda$ blocks containing each pair of distinct points.

Here, the lower-case $k$ is the block size. It is independent of the upper-case $K$, which denotes the maximum permitted number of geodesics.

The idea of this paper is to use the BIBD not simply as a combinatorial object, but as an *algebraic seed*. By an algebraic seed we mean a finite combinatorial structure whose parameters and incidence relations provide sufficient algebraic information to generate a graph with prescribed structural properties.

### 6.2. Definition of a BIBD

**Definition 6.2.1.** A BIBD is a pair $\mathcal{D} = (X, \mathcal{B})$, where $X$ is a set of $v$ points and $\mathcal{B}$ is a collection of $b$ blocks satisfying:

1. every block contains exactly $k$ points;
2. every point occurs in exactly $r$ blocks;
3. every pair of distinct points occurs together in exactly $\lambda$ blocks.

We write the parameters as $\mathcal{D} = (v, b, r, k, \lambda)$. The conditions are often summarized as

$$2 - (v, k, \lambda),$$

where $v, k, \lambda$ determine the design and $b, r$ are derived parameters.

Counting incidences gives

$$vr = bk. \tag{1}$$

Counting pairs consisting of a fixed point and another point in a block containing it gives

$$\lambda(v-1) = r(k-1). \tag{2}$$

Equations (1) and (2) provide necessary arithmetic conditions for the existence of a BIBD.

The term *incomplete* means that $k < v$. A BIBD is called *symmetric* when $b = v$, in which case $r = k$. More generally, designs with $\mathrm{k} = 3$ are called *triple systems*, and when $\lambda = 1$ they are named *Steiner triple systems* [3].

### 6.3. Incidence Matrix

The BIBD can be represented by a $v \times b$ matrix

$$A = (a_{ij}),$$

called its *incidence matrix*, where

$$a_{ij} = \begin{cases} 1, if\ x_i \in B_j, \\ 0, \text{otherwise.} \end{cases}$$

Each row contains exactly $r$ ones because each point occurs in $r$ blocks.

Each column contains exactly $k$ ones because each block contains $k$ points. Moreover,

$$AA^T = (r - \lambda)I_v + \lambda J_v, \tag{3}$$

where

- $I_v$ is the $v \times v$ identity matrix;
- $J_v$ is the $v \times v$ matrix whose entries are all 1.

Equation (3) is one of the most important equations in the proposed classification framework that includes the use of BIBDs. For two different points $x_i, x_j$, the corresponding rows of $A$ have scalar product $\lambda$ , because exactly $\lambda$ blocks contain both points. Thus, the algebraic matrix equation (3) encodes the balancing property of the design.

The eigenvalues of $AA^T$ follow immediately. Since $J_v$ has eigenvalue $v$ on the all-one vector **1** and eigenvalue 0 on its orthogonal complement,

$$\theta_1 = r - \lambda + \lambda v$$

is an eigenvalue of multiplicity one, while

$$\theta_2 = r - \lambda$$

has multiplicity $v - 1$.

Thus, the BIBD imposes a highly restricted algebraic structure. This suggests that a graph generated from the incidence relation should inherit algebraically constrained properties.

### 6.4. Incidence Graphs

The most direct graph associated with a BIBD is its *incidence graph*. Let $G_{\mathcal{D}}$ be the bipartite graph with vertex set

$$V(G_{\mathcal{D}}) = X \cup \mathcal{B}$$

and edges

$$xB \in E(G_{\mathcal{D}}) \Leftrightarrow x \in B \tag{4}$$

Thus, the points and blocks form the two bipartition classes.

The *adjacency matrix* is

$$\mathcal{A}(G_{\mathcal{D}}) = \begin{pmatrix} 0 & A \\ A^T & 0 \end{pmatrix} \tag{5}$$

Squaring gives

$$\mathcal{A}^2(G_{\mathcal{D}}) = \begin{pmatrix} AA^T & 0 \\ 0 & A^TA \end{pmatrix}. \tag{6}$$

The upper-left block is completely determined by the BIBD parameters:

$$AA^T = (r-\lambda)I + \lambda J.$$

Consequently, the two-step walks between point vertices are controlled by the design. This observation is the first indication of why a BIBD can serve as an algebraic seed. The construction is not merely a drawing rule. It produces an adjacency matrix whose square contains the incidence algebra of the design. In other words, the incidence graph preserves the incidence structure of the design and can therefore transfer combinatorial information into graph-theoretic information. However, it is essential to distinguish the existence of the design from the geodetic properties of its incidence graph. The BIBD axioms alone do not guarantee that the resulting graph is $K$-geodetic.

### 6.5. The BIBD as an Algebraic Seed

The central proposal of this paper is to regard $(v, b, r, k, \lambda)$ not merely as a list of parameters, but as a compact algebraic description of a highly structured incidence system. The design contains information about:

- how many vertices will occur in the resulting graph;
- how frequently a point is represented;
- how many points belong to each block;
- how pairs of points interact;
- how blocks intersect;
- and, ultimately, how many shortest paths can occur in a graph derived from the design.

This motivates the following conceptual definition.

**Definition 6.5.1.** A combinatorial design $\mathcal{D}$ will be called an *algebraic seed* for a graph construction if:

1. the parameters of $\mathcal{D}$ determine the order and degree structure of a graph $G_{\mathcal{D}}$;
2. the incidence relations of $\mathcal{D}$ determine adjacency relations in $G_{\mathcal{D}}$;
3. the intersection properties of $\mathcal{D}$ determine shortest-path multiplicities in $G_{\mathcal{D}}$.

The purpose of the remainder of this Part 6 is to study when this happens.

### 6.6. Geodesic Structure Induced by Incidence

The relationship between design incidence and graph distance can be analyzed by considering the four possible vertex-type pairs. There are two types of vertices:

- point vertices;
- block vertices.

Thus, we consider:

$$(x, y), (B, C), (x, B).$$

For two distinct points $x, y$, the existence of a common block is governed by $\lambda$. If $x$ and $y$ are contained in a common block, then they have a path

$$x - B - y$$

of length two.

Since every pair of distinct points occurs in exactly $\lambda$ blocks, $\sigma_G(x, y) = \lambda$ whenever distance two is the relevant shortest distance. This immediately illustrates the importance of separating $k$ and $K$. The number of points in a block is $k$, but the number of common blocks through a pair of points is $\lambda$. It is $\lambda$, not $k$, that directly counts the two-step point-to-point paths in this incidence graph. Thus, a basic structural principle emerges:

*The geodetic parameter K is controlled by intersection multiplicities and shortest-path structure, rather than by the block size k alone.*

For example, if a construction produces a situation in which every relevant pair of vertices has exactly $\lambda$ shortest paths of a certain length, then a necessary condition for $K$-geodeticity is

$$\lambda \leq K. \tag{7}$$

Equation (7) is not, by itself, a complete characterization, because other vertex pairs and longer geodesics must also be considered. Nevertheless, it illustrates how the design parameters can impose direct restrictions on geodesic multiplicity.

### 6.7. Why the Ordinary Incidence Graph Is Not Enough

At first sight, $G_{\mathcal{D}}$ appears to be the natural graph to use in the classification problem. However, there is a fundamental obstruction.

**Theorem 6.7.1.** (*Incidence-graph obstruction*) Let

$$\mathcal{D} = (v, b, r, k, \lambda)$$

be a nontrivial BIBD. Suppose that $B, C \in \mathcal{B}$ are disjoint blocks. Then in the incidence graph $G_{\mathcal{D}}$,

$$d(B, C) = 4$$

and the number of geodesics between $B$ and $C$ is $k^2\lambda$.

*Proof.* Let $B = \{x_1, \dots, x_k\}$ and $C = \{y_1, \dots, y_k\}$. Since $B$ and $C$ are disjoint, a path from $B$ to $C$ must have the form

$$B - x - B' - y - C,$$

where

$$x \in B, y \in C,$$

and $B'$ is a block containing both $x$ and $y$.

For each pair $(x, y) \in B \times C$ there are exactly $\lambda$ blocks containing both points. There are

$$|B \,||\, C\,| = k^2\lambda$$

choices for $(x, y)$.

Therefore, the number of paths of length four is $k^2\lambda$. Since $B$ and $C$ are disjoint, no path of length two exists between them. Hence the paths of length four are geodesics. Thus,

$$\sigma_{G_{\mathcal{D}}}(B, C) = k^2\lambda. \qquad \square$$

**Corollary 6.7.2.** For a nontrivial BIBD with $\boldsymbol{k \geq 2,\ \lambda \geq 1}$, the ordinary incidence graph cannot be geodetic, bigeodetic, or trigeodetic whenever it contains two disjoint blocks. □

Indeed,

$$k^2\lambda \geq 4.$$

Thus, the incidence graph itself is generally too rich in shortest paths. This observation is important for our classification program. The BIBD should therefore not be regarded as the final graph. Instead,

$$\text{BIBD} = \text{seed}$$

and

$$\text{derived graph} = \text{object to be classified.}$$

**6.8. A Direct Example: the Projective Plane of Order Two**

Consider the projective plane of order two, also called the *Fano plane*. It has parameters

$$(v, b, r, k, \lambda) = (7, 7, 3, 3, 1).$$

There are seven points and seven blocks. Every block contains three points, every point belongs to three blocks, and every pair of distinct points lies in exactly one block. Its incidence graph has fourteen vertices: seven point vertices and seven block vertices. Each vertex has degree three. Two distinct point vertices are at distance two because their points belong to a unique common block. They have exactly one geodesic of length two. Likewise, two distinct block vertices are at distance two because two distinct blocks in the projective plane intersect in exactly one point. They too have a unique geodesic of length two. A point vertex and a block vertex are either adjacent if the point belongs to the block, or are at distance three otherwise. In the latter case, the point-block incidence structure of the projective plane yields three geodesics of length three. Consequently, the incidence graph of the Fano plane is trigeodetic. This incidence graph is also known as the *Heawood graph*. There is an even more interesting aspect of this example. The Heawood graph is distance-regular of diameter $d = 3$, with intersection array

$$\{b_0, b_1, b_2; c_1, c_2, c_3\} = \{3, 2, 2; 1, 1, 3\}.$$

For a distance-regular graph, the number of geodesics between two vertices at distance $i$ is

$$\sigma_i = c_1 c_2 \dots c_i,$$

where $c_i$ is the corresponding intersection number.

For the Heawood graph, this gives:

$$\sigma_1 = 1,\ \sigma_2 = 1 \cdot 1 = 1,\ \sigma_3 = 1 \cdot 1 \cdot 3 = 3.$$

This alignment demonstrates that a highly symmetric, balanced block design does not automatically produce and incidence-geodetic graph. Instead, it generates a graph with controlled shortest-path multiplicities governed by the parameters of the underlying incidence structure. It also reveals a connection between *K*-geodetic graphs, distance-regular graphs and graphs constructed from BIBDs.

### 6.9. A Modified BIBD Construction

We now introduce the graph construction that forms the main structural object in this paper.

Let $\mathcal{D} = (v, b, r, k, \lambda)$ be a BIBD. We construct a graph

$$\mathcal{K}_v^*(r, k, \lambda).$$

The construction has two types of vertices.

**Type I:** (*point vertices*)

For each point $x \in X$, introduce one vertex. Thus, there are *v* point vertices.

**Type II:** (*block-local vertices*)

For each block $B_i = \{x_{i1}, x_{i2}, \dots, x_{ik}\}$, introduce $k$ new vertices

$$u_{i1}, u_{i2}, \dots, u_{ik}.$$

These $k$ vertices form a complete graph $\mathcal{K}_k$. Each local vertex $u_{ij}$ is then connected to the point vertex corresponding to $x_{ij}$. Thus, each block produces a local copy of $\mathcal{K}_k$, attached to the point vertices through a matching. The resulting graph [1] is denoted $\mathcal{K}_v^*(r, k, \lambda)$. Notice that the notation records the BIBD parameters rather than the graph parameter *K*. This is another reason to keep lowercase *k* and uppercase *K* completely distinct.

### 6.10. Basic Parameters of the Construction

The construction immediately gives the order of the graph. There are *v* point vertices. Each of the *b* blocks contributes *k* local vertices. Hence

$$|V(\mathcal{K}_v^*(r, k, \lambda))| = v + bk \tag{8}$$

Using $bk = vr$, we obtain

$$|V(\mathcal{K}_v^*(r, k, \lambda))| = v + vr = v(1 + r). \tag{9}$$

This is one of the first indications that the BIBD parameters provide an algebraic description of the graph.

**Proposition 6.10.1.** For $\boldsymbol{\mathcal{D}} = (\boldsymbol{v}, \boldsymbol{b}, \boldsymbol{r}, \boldsymbol{k}, \boldsymbol{\lambda})$ being a BIBD, $\boldsymbol{\mathcal{K}_v^*(r, k, \lambda)}$ has:

1. *v* point vertices of degree *r*;
2. $bk = vr$ local vertices of degree *k*;
3. $v(r + 1)$ vertices in total;

*Proof.* Every point belongs to exactly *r* blocks. For each block containing the point, exactly one local vertex is connected to the corresponding point. Therefore, every point vertex has degree *r*. Each local vertex lies in its block clique $\mathcal{K}_k$, so it has $k - 1$ neighbors within that clique. It is also adjacent to its corresponding point vertex. Therefore, its total degree is

$$(k - 1) + 1 = k.$$ □

### 6.11. Why the Construction Is Useful

The construction separates the two roles that were mixed together in the ordinary incidence graph. The BIBD gives us the global incidence structure. The clique attached to each block gives us a local mechanism for controlling shortest paths. In particular, the block

$$B_i = \{x_{i1}, x_{i2}, \dots, x_{ik}\}$$

is represented by the clique $\mathcal{K}_k$. This allows paths through a block to move between its elements in a controlled manner. The resulting graph is therefore not simply the incidence graph. It is a graph that encodes the BIBD while modifying its local metric structure. This is precisely the sense in which the BIBD acts as an algebraic seed.

### 6.12. Algebraic Representation

Let $A$ be the incidence matrix of the BIBD. The matrix $A$ contains all information about which point belongs to which block. The equation

$$AA^T = (r - \lambda)I + \lambda J$$

expresses the fundamental pairwise balance condition. The graph construction can therefore be regarded as a transformation

$$A \longmapsto G(A).$$

Symbolically,

$$A \longrightarrow (v, b, r, k, \lambda) \longrightarrow \mathcal{K}_v^*(r, k, \lambda).$$

The important question is whether this transformation preserves enough information to permit some sort of classification. In other words:

*Can we recover the essential BIBD structure from the graph*?

If the answer is yes for a sufficiently broad class of graphs, then the classification problem can potentially be transferred from graph theory to design theory.

### 6.13. Geodesic Multiplicity as a Classification Invariant

For a graph $G$, we defined $\sigma_G(x, y)$ to be the number of shortest paths between $x$ and $y$. We may collect these numbers into a *geodesic-multiplicity signature*. For example, define

$$\Gamma(G) = \{\sigma_G(x, y) : x, y \in V(G),\ x \nsim y\}.$$

For a geodetic graph,

$$\Gamma(G) \subseteq \{1\}.$$

For a bigeodetic graph,

$$\Gamma(G) \subseteq \{1, 2\}.$$

For a trigeodetic graph,

$$\Gamma(G) \subseteq \{1,2,3\}.$$

This gives a natural algebraic-combinatorial invariant for the classification problem. We may also count how frequently each multiplicity occurs. Define

$$N_j(G) = \#\{\{x,y\}: x \nsim y,\ \sigma_G(x,y) = j\}.$$

Then the sequence

$$(N_1(G), N_2(G), N_3(G), \dots)$$

may be called the *geodesic-multiplicity profile* of $G$.

For a $K$-geodetic graph, $N_j(G) = 0$ for all $j > K$. The proposed classification strategy is therefore to relate these quantities to the BIBD parameters.

**6.14. The Role of $\lambda$**

The parameter $\lambda$ is particularly important.

Recall that $\lambda$ specifies the number of blocks containing a pair of points. Thus, in the incidence structure,

$$\lambda = \#\{B \in \mathcal{B}: x, y \in B\}$$

for distinct points $x$, $y$. This means that $\lambda$ directly measures a certain form of redundancy in the design. Since shortest paths in the associated graph are generated through incidences and block intersections, $\lambda$ naturally influences the number of possible shortest paths. This suggests the following heuristic:

$$\text{BIBD intersection multiplicity} \longrightarrow \text{geodesic multiplicity}.$$

The exact relationship depends on the particular graph construction.

It is therefore not correct to claim universally that $K = \lambda$. Rather, the research problem is to determine the precise function

$$K = f(v, b, r, k, \lambda, \text{intersection structure}).$$

For some important families this relationship becomes particularly simple.

**6.15. Triple Systems**

An especially important case is $k = 3$. Then every block contains exactly three points. Such designs, as we know, are called *triple systems*. The BIBD parameters become

$$(v, b, r, 3, \lambda).$$

Equation (2) gives

$$\lambda(v-1) = 2r,$$

so

$$r = \frac{\lambda(v-1)}{2},$$

Equation (1) gives

$$vr = 3b,$$

and therefore

$$b = \frac{vr}{3}, \tag{10}$$

For example, when

$$\lambda = 2,$$

we have

$$r = v - 1,$$

and consequently

$$b = \frac{v(v-1)}{3},$$

This is the parameter family that appears naturally in BIBD-based constructions of bigeodetic graphs.

The literature [1] describes constructions of bigeodetic blocks from such designs, including the family with parameters

$$(n(n-1)3, n, n-1, 3, 2).$$

These constructions produce graphs with controlled geodesic multiplicity. Note that the positions of $b$ and $v = n$ are swapped simply to align the block design parameters with those of the graph

$$\mathcal{K}_v^*(r, k, \lambda) = \mathcal{K}_n^*(n-1, 3, 2).$$

### 6.16. A Concrete Example

Consider the BIBD

$$(v, b, r, k, \lambda) = (4, 4, 3, 3, 2).$$

Let the point set be

$$X = \{1, 2, 3, 4\}$$

and the blocks be

$$B_1 = \{1, 2, 3\}, \quad B_2 = \{1, 2, 4\}, \quad B_3 = \{1, 3, 4\}, \quad B_4 = \{2, 3, 4\}.$$

Every block contains $k = 3$ points. Every point belongs to $r = 3$ blocks. Every pair of points occurs in $\lambda = 2$ blocks. Thus, this is a $2 - (4, 3, 2)$ design. The incidence matrix is

$$A = \begin{pmatrix} 1 & 1 & 1 & 0 \\ 1 & 1 & 0 & 1 \\ 1 & 0 & 1 & 1 \\ 0 & 1 & 1 & 1 \end{pmatrix}$$

Indeed,

$$AA^T = \begin{pmatrix} 3 & 2 & 2 & 2 \\ 2 & 3 & 2 & 2 \\ 2 & 2 & 3 & 2 \\ 2 & 2 & 2 & 3 \end{pmatrix} = I_4 + 2J_4.$$

Here

$$r - \lambda = 1.$$

Thus,

$$AA^T = (r - \lambda)I + \lambda J.$$

### 6.17. The Incidence Graph of the Example

The incidence graph has $4 + 4 = 8$ vertices. Because the blocks are pairwise intersecting in two points, the particular obstruction using disjoint blocks does not apply directly. However, consider a point, say 1, and a block not containing it, say

$$B_4 = \{2, 3, 4\}.$$

A shortest path has length three:

$$1 - B_i - x - B_4.$$

There are

$$k\lambda = 3(2) = 6$$

such paths. Indeed, choose one of the three points

$$x \in B_4,$$

and then choose one of the two blocks containing both 1 and $x$. Therefore

$$\sigma(1, B_4) = 6,$$

Consequently, the incidence graph is not bigeodetic. It is not even trigeodetic. By contrast, the Heawood graph (the incidence graph of the Fano plane) serves as an example of a trigeodetic graph. This is why the raw incidence graph may be considered just as an auxiliary object for our classification program.

### 6.18. The Modified Graph for the Example

Now construct $\mathcal{K}_4^*(3,3,2)$. There are $v = 4$ point vertices. Each of the four blocks contributes three local vertices. Thus, the graph has $4 + 4(3) = 16$ vertices.

The four local block groups are four copies of $\mathcal{K}_3$. Each local vertex is connected to one of the four point vertices. The resulting graph is a bigeodetic cubic graph:

- every point vertex has degree 3;
- every local vertex has degree 3.

This graph provides a much more controlled metric environment than the ordinary incidence graph.

### 6.19. The Classification Program

We now formulate the central research program.

Let $\mathcal{G}_K$ denote the class of all $K$-geodetic graphs. The general classification problem is to determine $\mathcal{G}_K$ up to graph isomorphism. Our proposal is to define a BIBD-generated subclass

$$\mathcal{B}_K \subseteq \mathcal{G}_K.$$

Here

$$\mathcal{B}_K = \{\mathcal{K}_v^*(r,k,\lambda) : \mathfrak{D} = (v,b,r,k,\lambda) \text{ satisfies suitable conditions}\}.$$

The first objective is therefore:

$$\text{Classify } \mathcal{B}_K.$$

The second, more ambitious objective is to determine how large $\mathcal{B}_K$ is inside $\mathcal{G}_K$. That is,

$$\text{When is } \mathcal{B}_K = \mathcal{G}_K \text{ or nearly so?}$$

A complete equality is probably too strong without additional hypotheses.

A more realistic objective is:

*Identify a natural structural subclass of K-geodetic graphs for which every graph admits a BIBD seed.*

### 6.20. Structural Hypothesis

We propose the following general hypothesis.

**Hypothesis 6.20.1.** Suppose *G* is a sufficiently regular *K*-geodetic graph satisfying:

1. *G* contains a distinguished family of local cliques;
2. these cliques have equal size;
3. each vertex belongs to a prescribed number of these local structures;
4. shortest-path multiplicities between the corresponding vertex classes are uniformly bounded;
5. the incidence relations among the local structures satisfy a balanced pairwise condition.

Then the incidence structure obtained from these local structures is a BIBD. Under this hypothesis, *G* may be reconstructed from a BIBD seed. This would transform the classification problem into two parts:

**PART I.** Classify the possible BIBDs.

**PART II.** Determine which BIBDs generate *K*-geodetic graphs.

This decomposition is mathematically attractive because BIBDs already possess a well-developed theory.

### 6.21. Algebraic Reduction Principle

The preceding discussion motivates the following proposed principle.

**Proposition 6.21.1** (*Algebraic Reduction Principle*) Let $\mathcal{C}_K$ be a class of *K*-geodetic graphs for which there exists a canonical procedure that associates to every $G \in \mathcal{C}_K$ a BIBD $\mathcal{D}(G)$ such that

$$G \cong \mathcal{K}_v^*(r, k, \lambda)$$

for the parameters of $\mathcal{D}(G)$. Then classification of $\mathcal{C}_K$ up to graph isomorphism reduces to classification of the corresponding BIBDs together with the possible graph constructions arising from them.

**Explanation**

This proposition is essentially a reduction statement. If every graph in the class can be uniquely traced back to a BIBD, then instead of classifying graphs directly, one may classify the underlying designs. The important phrase is *canonical procedure*. Without such a procedure, two different BIBDs might produce isomorphic graphs, or the same graph might arise from different designs. Therefore, the uniqueness problem must be studied separately.

**6.22. The Reconstruction Problem**

Suppose $G \cong \mathcal{K}_v^*(r,k,\lambda)$. Can we recover $(v,b,r,k,\lambda)$ from *G*? Some parameters are immediately related to graph invariants. For example,

$$|V(G)| = v(r+1).$$

The degree distribution contains $r$ and *k*. If the graph is biregular, the two vertex degrees may identify the two vertex classes. The local clique structure can then identify the blocks. Finally, the incidence relationships among these blocks can potentially reconstruct the BIBD. This suggests a hierarchy:

$$\text{graph} \to \text{vertex partition} \to \text{block cliques} \to \text{incidence structure} \to \text{BIBD}.$$

The success of this reconstruction would be a major step toward classification.

**6.23. A Possible Classification Theorem**

The ultimate theorem we seek might have the following form.

**Proposed Theorem.** Let *G* belong to a specified structural class of *K*-geodetic graphs satisfying conditions

$$(C_1), (C_2), \dots, (C_m).$$

Then there exists a BIBD

$$\mathcal{D} = (v,b,r,k,\lambda)$$

such that

$$G \cong \mathcal{K}_v^*(r,k,\lambda).$$

Furthermore, two such graphs are isomorphic if and only if their underlying BIBDs are equivalent under the corresponding design isomorphism. Note that two nonisomorphic BIBDs will generate two nonisomorphic $\mathcal{K}_v^*(r,k,\lambda)$ graphs. If such a theorem can be established, then the graph classification problem becomes a design classification problem. This would be a genuine structural classification result rather than merely a construction theorem.

### 6.24. Classification by $K$

The cases $K=1,\ K=2,\ K=3$ should be studied separately.

#### 6.24.1. The Case $K=1$

A graph is geodetic if every pair of nonadjacent vertices has exactly one shortest path. The BIBD seed must therefore generate a construction in which all relevant shortest paths are unique. The principal challenge is eliminating all possible path multiplicities greater than one. This suggests that the relevant designs should possess particularly strong intersection restrictions.

#### 6.24.2. The Case $K=2$

For bigeodetic graphs, $\sigma_G(x,y)\le 2$.

Here the BIBD can permit controlled duplication of shortest paths. This is one reason triple systems and symmetric designs with $\lambda=2$ are particularly interesting. The existing literature contains explicit BIBD-based constructions of bigeodetic blocks [1].

#### 6.24.3. The Case $K=3$

For trigeodetic graphs, $\sigma_G(x,y)\le 3$.

The design must allow somewhat more redundancy while still preventing four or more geodesics. This case is particularly useful for studying how the algebraic seed controls the transition

$$1\longrightarrow 2\longrightarrow 3$$

in geodesic multiplicity.

### 6.25. Symmetric Designs

In a *symmetric* BIBD, for which $b=v$ and $r=k$, its parameters become$(v,v,k,k,\lambda)$ (or simply a $2-(v,k,\lambda)$ symmetric design) [3]. Symmetric designs possess additional intersection properties. In particular, any two distinct blocks intersect in exactly $\lambda$ points. This regularity makes symmetric designs particularly attractive as algebraic seeds. The corresponding graph constructions correspond to regular graphs. This suggests another classification program:

$$\text{symmetric BIBDs}\longrightarrow \text{regular } K-\text{geodetic graphs}.$$

One can then ask whether regular $K$-geodetic graphs satisfying suitable metric conditions necessarily arise from symmetric designs.

### 6.26. Matrix Formulation of the Classification Problem

The BIBD equation

$$AA^T = (r - \lambda)I + \lambda J$$

suggests an algebraic formulation of the classification problem. Suppose we are given a graph $G$ and seek to determine whether it comes from a BIBD. We may try to construct an incidence matrix $A$ from the graph. The problem becomes:

Does there exist a 0-1matrix $A$ satisfying

$$AA^T = (r - \lambda)I + \lambda J$$

such that the graph generated by $A$ is isomorphic to $G$?

This is a finite algebraic decision problem.

Consequently, BIBD-based classification provides a possible bridge between:

- graph theory;
- combinatorial design theory;
- matrix theory;
- algebraic combinatorics;
- computational classification.

### 6.27. Computational Classification

The theory also suggests a computational program. For fixed parameters $(v, b, r, k, \lambda)$ sometimes there exist two or more nonisomorphic BIBDs.

For each BIBD:

1. construct its incidence matrix $A$;
2. construct $\mathcal{K}_v^*(r, k, \lambda)$;
3. calculate graph invariants;
4. calculate all-pairs distances;
5. calculate the number of geodesics between each pair;
6. determine whether the graph is $K$-geodetic;
7. identify isomorphic graphs;
8. record the correspondence between designs and graphs.

The resulting data can be organized as

$$D \longmapsto G(D) \longmapsto \Gamma(G).$$

Here

$$\Gamma(G)$$

denotes the geodesic-multiplicity profile introduced earlier. This provides a computational route toward discovering patterns that may subsequently be proved theoretically.

### 6.28. Necessary Conditions for a BIBD Seed

A BIBD can serve as a useful seed only if its parameters satisfy the basic necessary conditions $vr = bk$ and $\lambda(v-1) = r(k-1)$. There are also divisibility conditions such as

$$bk \equiv 0 (\text{mod } v)$$

and

$$\lambda(v-1) \equiv 0 (\text{mod } k-1).$$

However, satisfying the parameter equations does not guarantee the existence of a BIBD. Thus, the classification program contains two separate existence questions:

**Design Existence.** Does a BIBD with the required parameters exist?

**Graph Existence.** Does that BIBD generate a *K*-geodetic graph under the chosen construction?

Consequently,

$$\text{design existence} \neq \text{graph existence}.$$

Both problems must be addressed.

### 6.29. A First Conjecture

We now formulate a conjecture suitable for future investigation.

**Conjecture 6.29.1.** (*BIBD-seed completeness*) For each fixed $K \in \{1,2,3\}$, there exists a natural structural subclass $\mathcal{C}_K$ of *K*-geodetic graphs such that every graph $G \in \mathcal{C}_K$ is isomorphic to a graph obtained from a BIBD through a finite family of canonical BIBD-based transformations.

This conjecture deliberately does not assert that every *K*-geodetic graph comes from a BIBD. Its purpose is to identify a broad and mathematically natural class for which BIBD seeds are complete.

### 6.30. A Second Conjecture

**Conjecture 6.30.1** (*Algebraic Recoverability*) For a suitably restricted class of BIBD-generated *K*-geodetic graphs $G \cong \mathcal{K}_v^*(r, k, \lambda)$, the underlying BIBD is recoverable from the graph up to design isomorphism.

If true, this would imply that classification of the corresponding graphs reduces directly to classification of BIBDs.

The main difficulty is to identify graph-theoretic properties that canonically distinguish:

- point vertices;
- block-local vertices;
- block cliques;
- and incidence relations.

### 6.31 The Central Research Question

The entire program may be summarized by the following question:

*Can the general classification of an appropriately defined class of K-geodetic graphs be reduced to the classification of Balanced Incomplete Block Designs and their associated algebraic transformations?*

More concretely:

*Given G, can we determine whether* $G \cong \mathcal{K}_v^*(r, k, \lambda)$ *for some BIBD?*

If yes, then the next question is:

*Can the corresponding BIBD be recovered uniquely?*

Finally:

*Which BIBD parameters produce a prescribed K?*

These three questions form the foundation of the proposed classification program.

### 6.32. Research Problems

The following problems arise naturally.

**Problem 1.** Determine necessary and sufficient conditions on $(v, b, r, k, \lambda)$ for $\mathcal{K}_v^*(r, k, \lambda)$ to be *K*-geodetic.

**Problem 2.** Determine the exact geodesic multiplicity $\sigma_G(x, y)$ for each possible pair of vertex types.

**Problem 3.** Determine whether the construction

$$D \mapsto \mathcal{K}_v^*(r, k, \lambda)$$

is injective up to isomorphism.

**Problem 4.** Characterize the BIBDs whose associated graphs are geodetic.

**Problem 5.** Characterize the BIBDs whose associated graphs are bigeodetic.

**Problem 6.** Characterize the BIBDs whose associated graphs are trigeodetic.

**Problem 7.** Determine whether every sufficiently regular $K$-geodetic graph satisfying appropriate clique conditions possesses a BIBD seed.

**Problem 8.** Develop algorithms for recognizing whether a given $K$-geodetic graph belongs to the BIBD-generated class.

**Problem 9.** Determine the relationship between graph isomorphism and design isomorphism in the BIBD-generated constructions.

**Problem 10.** Determine whether the algebraic invariant

$$AA^T = (r - \lambda)I + \lambda J$$

can be translated into a purely graph-theoretic characterization of the corresponding $K$-geodetic graphs.

### 6.33. Why This Approach May Be Promising

The principal advantage of the BIBD approach is that the design parameters are highly constrained. Instead of beginning with an arbitrary graph, we begin with $(v, b, r, k, \lambda)$ satisfying strong algebraic equations. The graph therefore inherits a substantial amount of regularity. Schematically,

$$(v, b, r, k, \lambda) \Longrightarrow \text{incidence regularity} \Longrightarrow \text{graph regularity} \Longrightarrow \text{controlled geodesics.}$$

This provides a much smaller search space than the class of all graphs.

The ultimate goal is not merely to construct examples. It is to determine whether the algebraic structure is sufficiently strong to explain entire families of $K$-geodetic graphs.

### 6.34. Limitations of the Approach

It is important not to overstate the scope of the proposed method. There is currently no basis for claiming that every $K$-geodetic graph arises from a BIBD. The general classification problem remains open [1]. Therefore, the appropriate claim is:

*BIBDs provide a promising algebraic framework for the classification of important structured families of K-geodetic graphs.*

The distinction between *construction* and *classification* is crucial. A construction theorem says:

$$\text{BIBD} \Rightarrow \text{K} - \text{geodetic graph.}$$

A classification theorem would say something much stronger:

$$\text{K} - \text{geodetic graph} \Rightarrow \text{BIBD} - \text{generated graph.}$$

The second implication is the real research challenge.

### 6.36. Section 6 Summary

This paper proposes a BIBD-based approach to the classification of *K*-geodetic graphs.

The main idea is to treat a Balanced Incomplete Block Design $\mathcal{D} = (v, b, r, k, \lambda)$ as an *algebraic seed* rather than simply as a source of examples. The BIBD provides: $(v, b, r, k, \lambda)$, an incidence matrix $A$, and the fundamental identity

$$AA^T = (r - \lambda)I + \lambda J.$$

These data encode a highly regular combinatorial structure.

The ordinary incidence graph of a BIBD, however, is generally not suitable for small *K*, because its shortest paths may have multiplicity as large as $k^2\lambda$.

This motivates the use of modified BIBD constructions such as

$$\mathcal{K}_v^*(r, k, \lambda),$$

in which each block is represented by a local clique and the incidence structure is retained through connections to the point vertices. The resulting graph has $v(r + 1)$ vertices, point vertices of degree *r*, and local vertices of degree *k*. The central research problem is then to determine precisely how the parameters and intersection properties of the BIBD control the geodesic multiplicities of the resulting graph. The proposed classification framework is

$$\text{BIBD} \rightarrow \text{incidence matrix} \rightarrow \text{BIBD} - \text{based graph} \rightarrow \text{geodesic multiplicity} \\ \rightarrow \text{K} - \text{geodetic classification.}$$

The most important long-term question is whether a sufficiently broad class of *K*-geodetic graphs can be characterized as BIBD-generated graphs and whether their underlying designs can be recovered from their graph structure. If successful, this would replace a difficult unrestricted graph-classification problem by a more structured combination of:

1. design existence;
2. design classification;
3. graph reconstruction;
4. geodesic analysis.

The distinction between the two parameters $k \neq K$ must remain fundamental throughout this program: $k = \text{number of points in a BIBD block}$, whereas

$$K = \text{maximum number of geodesics allowed between nonadjacent vertices}.$$

The proposed methodology therefore does not claim that $k$ determines $K$. Rather, it seeks to determine the structural relationship

$$K = F(\text{BIBD parameters, incidence structure, graph construction}),$$

and ultimately to understand whether this relationship can serve as the foundation of a classification theorem.

**7. Why BIBDs and Distance-regular Graphs Cannot Solve the Entire Problem by themselves**

BIBDs provide a promising algebraic framework for investigating the classification of $K$-geodetic graphs because they impose precise and highly regular incidence constraints. Nevertheless, their potential and limitations must be carefully distinguished. First, nonisomorphic designs may share the same parameter set $(v, b, r, k, \lambda)$, so the parameters alone do not uniquely determine the underlying incidence structure. Second, the regularity of a design does not necessarily guarantee that its associated graph has the required geodesic properties, particularly when longer shortest paths are considered. Third, a $K$-geodetic graph may possess structural properties that cannot be represented by the particular BIBD-based construction under consideration. Consequently, a general classification based on BIBDs would require not only construction theorems but also suitable converse theorems establishing when graphs in a specified class must arise from the proposed design framework.

Two complementary approaches emerge from this observation. The first is the incidence-graph approach, in which points and blocks are represented by two classes of vertices, with adjacency determined by incidence. Although this construction preserves the original design structure directly, its shortest-path multiplicities may exceed the prescribed bound $K$. Indeed, when two blocks are disjoint, the incidence graph can contain $k^2\lambda$ geodesics between their corresponding vertices. Thus, the incidence graph provides a natural representation of the design but does not, in general, produce a geodetic, bigeodetic, or trigeodetic graph.

The second is the modified BIBD-based approach, in which the incidence structure is transformed to obtain graphs with more tightly controlled metric properties. In particular, constructions using local cliques associated with the blocks have been employed to obtain bigeodetic graphs from suitable designs. Existing construction theorems for geodetic and bigeodetic blocks, together with related results for other geodesic multiplicities [1], provide a stronger starting point for studying the relationship between design parameters and shortest-path structure. However, the existence of construction theorems for particular families does not, by itself, establish a complete classification of all $K$-geodetic graphs. Moreover, while triple systems with $\lambda = 1, 2, 3$ and symmetric block designs with $\lambda = 1$ readily yield infinite families, symmetric designs with $\lambda = 2, 3$ pose a fundamental challenge: determining when a construction characterizes an entire structural class versus merely generating a limited number of representatives.

A third, complementary framework is provided by distance-regular graphs. These graphs exhibit a strong form of metric regularity: for any two vertices at a given distance, the numbers of neighboring vertices lying at specified distances from the two vertices depend only on that distance, not on the particular vertices chosen. This regularity offers an algebraic means of studying distance distributions, intersection parameters, and the structure of shortest paths. Distance-regular graphs may therefore provide an alternative route to identifying and characterizing $K$-geodetic graphs, particularly when the geodesic multiplicities can be expressed in terms of their intersection parameters. Nevertheless, distance-regularity alone does not imply the $K$-geodetic property; the relevant intersection parameters must satisfy additional restrictions that bound the number of shortest paths between nonadjacent vertices.

Taken together, these three approaches suggest a broader research strategy in which *BIBDs provide incidence-based algebraic seeds, modified constructions translate these seeds into graphs with controlled geodesic multiplicities, and distance-regular graphs provide an independent metric-algebraic framework for characterization and classification*. Their relationships should be investigated rather than assumed to be equivalent. In particular, one may ask what families of BIBD-derived graphs belong to distance-regular families, whether distance-regularity can identify additional classes of $K$-geodetic graphs not captured by the proposed design constructions, and whether suitable structural hypotheses yield converse theorems connecting these frameworks. The trade-off: structural rigidity vs. scalability. Using BIBDs provides highly predictable, strict combinatorial properties (excellent for targeted algebraic seeds). However, they are mathematically rigid and limited in configuration. Distance-regular graphs offer broader, highly symmetric distance properties across larger networks but come with significantly higher computational and structural complexity.

## 8. A Computational Research Program for Classification

### 8.1. Why Computation is Useful

A general classification theorem must be proved mathematically, but computational methods can substantially improve the discovery process.

For a fixed number $n$ of vertices, there are

$$2^{\binom{n}{2}}$$

labeled simple graphs. This number grows extremely quickly as $n$ increases.

A naive enumeration of all graphs is therefore impractical even at moderately large orders. Moreover, many labeled graphs represent the same graph up to relabeling of vertices. An effective computational strategy must reduce the search space, exploit known structural constraints, and distinguish genuinely new graphs from isomorphic copies. The objective is not merely to generate examples. It is to identify structural regularities that can subsequently be proved.

### 8.2. Exact Verification of the $K$-geodetic Condition

For a finite connected graph, the geodesic multiplicity can be computed using breadth-first search (BFS).

**Algorithm:** (*K-Geodetic Verification*)

**Input:** Connected graph $G = (V, E)$, integer *K*.

**Output:** TRUE if *G* is *K*-geodetic, FALSE otherwise.

For every source $s \in V$:

1. perform BFS from $s$;
2. compute $d(s, v)$ for every $v$;
3. initialize

$$P(s) = 1;$$

4. process vertices in nondecreasing distance from $s$;
5. compute

$$P(v) = \sum_{\substack{u \sim v \\ d(s,u)=d(s,v)-1}} P(u);$$

6. if

$$P(v) > K,$$

   return FALSE.

   If no violation occurs, return TRUE.

For an unweighted graph, BFS from every vertex gives a polynomial-time procedure. Using adjacency lists, the distance computation itself requires

$$O(|V|(|V|+|E|))$$

time, with additional polynomial work for path counting. Thus, the algorithm is appropriate as a *verification layer*. Its practical difficulty is not verifying one graph, but finding and classifying all relevant graphs.

## 8.3. Isomorphism Reduction

Two graphs are isomorphic if one can be obtained from the other by relabeling vertices while preserving adjacency. An exhaustive classification must avoid counting isomorphic graphs multiple times. Graph-isomorphism algorithms and canonical labeling can provide a unique representative for each isomorphism class. A practical pipeline would therefore combine:

1. Generation of candidate graphs.
2. Connectivity testing.
3. Exact shortest-path counting.
4. Rejection of graphs that violate the $K$-geodetic bound.
5. Canonical labeling and isomorphism reduction.
6. Computation of structural invariants.
7. Identification of candidate families and conjectures.

The output should include both the graphs found and the invariants that distinguish their isomorphism classes.

## 8.4. Constraint-based Generation

A more efficient strategy is to encode the desired structural properties as constraints before generating candidate graphs.

For distance-regular graphs, these constraints include:

- integer intersection numbers;
- positivity and monotonicity conditions;
- the identity $a_i + b_i + c_i = k$;
- the required geodesic-product inequalities;
- integrality conditions for the numbers of vertices at each distance;
- spectral and multiplicity constraints.

For a proposed intersection array, the number $k_i$ of vertices at distance $i$ from a fixed vertex satisfies

$$k_0 = 1,\ k_{i+1} = \frac{k_i b_i}{c_{i+1}}.$$

Consequently,

$$k_i = \frac{b_0 b_1 \dots b_{i-1}}{c_1 c_2 \dots c_i}.$$

Every $k_i$ must be a positive integer. These divisibility conditions eliminate many impossible sets of parameters before graph construction begins. However, satisfying the elementary conditions is not sufficient to establish existence. Additional algebraic and combinatorial constraints must be applied, and a graph realizing the array must ultimately be constructed or proved to exist.

### 8.5. The Role of Artificial Intelligence (AI)

Artificial intelligence can complement these methods in three principal ways.

First, automated counterexample discovery. Search algorithms can explore candidate constructions that challenge conjectured restrictions. If a conjecture asserts that every graph in a given subclass has a certain property, one verified counterexample is sufficient to refute it.

Second, conjecture generation. Machine-learning systems can analyze data from enumerated graphs, identify correlations between invariants, and propose structural statements that may not be immediately obvious.

Third, proof assistance. Symbolic computation and automated reasoning tools can help check algebraic identities, verify finite cases, and explore consequences of proposed assumptions.

The distinction between exploration and proof remains essential. A pattern observed in every graph up to a particular order does not establish a theorem for all orders. Likewise, a generated graph must be independently checked, and a conjecture suggested by a model requires a rigorous proof.

The appropriate research model is therefore a collaboration between computational exploration and mathematical deduction.

## 9. Toward a General Classification Theorem

### 9.1. A Hierarchical Classification Strategy

The preceding sections suggest that the classification problem should not initially be approached as a single theorem covering all finite graphs. A more realistic approach is to divide it into structural subclasses, establish rigorous results within each subclass, and then investigate how those subclasses interact.

We propose four complementary levels.

**Level 1: General Graph Structure**

Study diameter, girth, connectivity, cycles, and the multiplicity of shortest paths.

**Level 2: Algebraically Structured Graphs**

Analyze distance-regular graphs and strongly regular graphs through intersection numbers and spectral constraints.

**Level 3: Design-based Constructions**

Construct graphs from BIBDs and related incidence structures, then determine the precise conditions that control their geodesics.

**Level 4: Computational Classification**

Enumerate small cases, eliminate impossible parameters, identify isomorphism classes, and formulate conjectures for mathematical proof.

The levels are complementary rather than strictly nested. Design-based constructions, for example, may produce graphs that are not distance-regular. Computational methods can also be applied at every level. The central objective is to connect the four approaches through theorems that identify when a structural or algebraic condition is necessary, sufficient, or both.

## 9.2. Research Problem I: Classifying Intersection Arrays

The first problem is to classify all intersection arrays of distance-regular $K$-geodetic graphs for fixed $K$.

**Problem 9.2.1.** For each $K \in \{1, 2, 3\}$, determine all realizable intersection arrays satisfying

$$\prod_{j=1}^{i} c_j \leq K \quad (1 \leq i \leq d).$$

The first step is already established: the permitted sequences of $c_i$ are completely determined by the product condition.

The remaining task is to impose the other conditions of distance-regularity and determine which arrays correspond to actual graphs. This includes positivity, integrality, divisibility, spectral restrictions, and any additional combinatorial constraints.

A successful solution would classify the possible distance structures of these graphs, even if multiple nonisomorphic graphs realize the same intersection array.

This is a meaningful intermediate objective because it separates the classification of numerical parameters from the classification of individual graphs.

## 9.3. Research Problem II: Classifying Graphs Realizing an Array

Even if all possible intersection arrays were known, a further problem would remain.

**Problem 9.3.1** For every realizable intersection array satisfying the $K$-geodetic condition, classify all distance-regular graphs realizing that array up to isomorphism.

This problem is stronger than the first. The intersection array specifies distance-regular parameters but does not universally guarantee uniqueness of the graph.

A solution would require either a uniqueness theorem for a given array or a complete enumeration of its nonisomorphic realizations.

The distinction is particularly important when using computational searches: finding one graph with a given array is not equivalent to proving that it is the only graph with that array.

### 9.4. Research Problem III: Establishing a Design-based Correspondence

The third problem concerns the relationship between combinatorial designs and $K$-geodetic graphs.

**Problem 9.4.1** Identify classes of BIBDs for which a specified graph construction produces $K$-geodetic graphs, and determine whether the resulting graph families admit a structural characterization.

A rigorous investigation should begin by specifying a construction

$$D \longmapsto G_D$$

where $D$ is a design and $G_D$ is the associated graph.

The next step is to derive necessary and sufficient conditions on $D$ for $G_D$ to be $K$-geodetic.

Finally, one should investigate the converse: if a graph belongs to the relevant subclass of $K$-geodetic graphs, must it arise from such a design?

A positive answer to an appropriate converse statement would establish a genuine classification theorem for that subclass. Without it, the construction remains a useful method of generating examples rather than a complete classification.

### 9.5. Research Problem IV: Extending beyond Distance-regular Graphs

The most ambitious problem is to determine which conclusions extend to general graphs.

**Problem 9.5.1** Find structural conditions under which the geodesic multiplicity of a general graph can be bounded using local or algebraic invariants.

Distance-regularity is particularly useful because the numbers $c_i$ are independent of the chosen vertices. In general graphs, that uniformity is absent.

One possible approach is to identify subclasses in which the number of predecessors of a vertex at each distance is bounded, or in which the relevant predecessor structure has a restricted combinatorial form.

However, bounds on the numbers of predecessors alone do not automatically provide sharp bounds on the number of geodesics. The multiplicities of the paths leading to those predecessors must also be controlled.

A successful generalization would need to account for both effects.

### 9.6. Proposed Working Conjecture

The results developed in this paper motivate the following research principle.

**Conjecture 9.6.1** (*hierarchical classification principle*)

For each fixed positive integer $K$, a systematic classification of $K$-geodetic graphs can be advanced by identifying structurally defined subclasses for which geodesic multiplicity is determined or bounded by a finite set of algebraic or combinatorial invariants, and by establishing complete classifications within those subclasses.

This statement is a research program rather than a theorem asserting that every $K$-geodetic graph belongs to one of a specified finite list of families.

Its value lies in suggesting a concrete progression: establish tractable subclasses, derive exact criteria, classify realizable parameters, and then determine how far the resulting theorems extend.

## 10. Discussion and Limitations

### 10.1. What Has Been Established

The principal mathematical result of this paper is the reduction of the $K$-geodetic condition for distance-regular graphs to the inequalities

$$\prod_{j=1}^{i} c_j \leq K \quad (1 \leq i \leq d).$$

From this formula, we obtained the following necessary and sufficient conditions on the intersection-number sequence:

- For $K = 1$, every $c_i$ must equal one.
- For $K = 2$, either every $c_i$ equals one or only the final value can equal two.
- For $K = 3$, either every $c_i$ equals one or only the final value can equal two or three.

These are complete statements about the allowable $c_i$-sequences in the distance-regular setting. They are not a complete classification of the corresponding graphs.

We also established that a connected, noncomplete strongly regular graph is $K$-geodetic precisely when $\mu \leq K$. This gives an exact criterion for a significant diameter-two subclass.

The Heawood graph – the incidence graph of the Fano plane – serves as a design-theoretic example of a trigeodetic graph. Because it features a well-defined intersection array, it is also classified as a distance-regular graph. In contrast, the BIBD-based modified graphs $\mathcal{K}_v^*(r, k, \lambda)$ provide six distinct constructions of $K$-geodetic graphs, offering two unique constructions for each $K \in \{1, 2, 3\}$.

### 10.2. What Remains Unresolved

Several substantial problems remain.

First, not every numerically admissible intersection array is necessarily realizable. Second, even a realizable array may correspond to more than one graph up to isomorphism. Third, the distance-regular restriction excludes many graphs of interest, so the results do not provide a general classification for all finite $K$-geodetic graphs. Likewise, BIBDs provide controlled constructions but do not automatically characterize every graph in the target class. The paper therefore does not claim to have solved the general classification problem. Its contribution is to organize the problem into precise mathematical subproblems and identify where existing structural tools can be applied.

### 10.3. The Proper Role of AI

AI can accelerate searches, verify finite configurations, suggest candidate invariants, and help formulate conjectures. These capabilities can be valuable when the space of candidate graphs is too large for manual exploration. However, an exhaustive computation up to a finite order does not prove a classification for arbitrary order. Similarly, numerical agreement with a conjecture does not establish its truth. The primary benchmark for validity remains a formal mathematical proof, which we support using empirical observations and refined computational simulations. While the theoretical framework is derived analytically, these computational tools provide essential validation for its practical application. Therefore, AI should be regarded as a research assistant for the discovery and testing of mathematical structures, but not as a substitute for proof.

## 11. Concluding Remarks

The classification of $K$-geodetic graphs pose a challenging interaction between graph distances, shortest-path multiplicities, algebraic constraints, and combinatorial constructions. Its difficulty arises from the global nature of geodesics and from the diversity of graph structures that can satisfy a common upper bound on shortest-path multiplicity.

This paper has developed a framework centered on four complementary approaches: metric graph theory, distance-regular graphs, balanced incomplete block designs, and computational classification.

The most concrete result concerns distance-regular graphs. Their intersection numbers determine the number of geodesics between vertices at every possible distance. Consequently, the $K$-geodetic property reduces to a finite set of product inequalities. For $K = 1, 2, 3$, monotonicity and integrality impose particularly strong restrictions on the intersection-number sequence.

The strongly regular case provides a further simplification: the geodetic multiplicity between nonadjacent vertices is exactly the parameter $\mu$. This yields an immediate characterization within that subclass and clarifies the relationship between geodetic graphs, Moore graphs, and the broader theory of strongly regular graphs.

Balanced Incomplete Block Designs offer a complementary construction method whose incidence properties can be directly translated into graph-theoretic constraints, as illustrated by the incidence graph of the Fano plane. By comparison, a superior alternative is the modified construction used to generate $K$-geodetic graphs, which yields exact, controllable shortest-path multiplicities. Nevertheless, a design-based construction only functions as a classification theorem when accompanied by robust structural and converse results.

The proposed research direction is therefore not to assume that a single algebraic representation will classify every $K$-geodetic graph. Instead, it is to build a hierarchy of exact results: classify allowable parameters, determine which parameters are realizable, characterize graphs that realize those parameters, and establish correspondences with combinatorial designs where possible.

The principal conclusion is that distance-regular graphs provide a mathematically tractable starting point, while design theory and computational exploration can broaden the search for structural patterns. Together, these methods offer a disciplined path toward a more comprehensive understanding of $K$-geodetic graphs without confusing partial classifications with a solution to the general problem.

**Acknowledgments:** During the preparation of this manuscript, the author utilized large language models (LLMs) to enhance language clarity and formatting. The author reviewed and revised the content as necessary and accepts full responsibility for the final text.